\pdfoutput=1

\documentclass[aps,prl,reprint,showpacs,superscriptaddress]{revtex4-1}

\usepackage{graphicx}
\usepackage{graphics}
\usepackage{amsmath}
\usepackage{amssymb}
\usepackage{amsfonts}
\usepackage{dcolumn}
\usepackage{dsfont}
\usepackage{latexsym}
\usepackage{rotating}
\usepackage{color}
\usepackage{latexsym}
\usepackage{bbm}
\usepackage{subfigure}
\usepackage{float}
\usepackage{epsfig}
\usepackage{epsf}
\usepackage{psfrag}
\usepackage{bm}
\usepackage{amsthm}
\usepackage{eucal}
\usepackage{mathrsfs}
\usepackage{url}
\usepackage{braket}
\usepackage{epstopdf}
\graphicspath{ {./images/} }
\newcommand{\RNum}[1]{\uppercase\expandafter{\romannumeral #1\relax}}
\usepackage{color} 

\usepackage{hyperref}
\hypersetup{
colorlinks=true,final=true,
        linkcolor=blue,
        citecolor=blue,
        filecolor=blue,
        urlcolor=blue,
}
  
\begin{document}
  
\title{ Two-dimensional Rare-earth Halide Based Single Phase Triferroic}
  
\author{Srishti Bhardwaj}
\affiliation{ Department of Physics, Indian Institute of Technology Roorkee, Roorkee - 247667, Uttarakhand, India}
\author{T. Maitra}
\affiliation{ Department of Physics, Indian Institute of Technology Roorkee, Roorkee - 247667, Uttarakhand, India}
\email{tulika.maitra@ph.iitr.ac.in}

\date{\today}

\begin{abstract}
Two-dimensional multiferroic materials are highly sought after due to their huge potential for applications in nanoelectronic and spintronic devices. Here, we predict, based on first-principle calculations, a single phase {\it triferroic} where three ferroic orders; ferromagnetism, ferroelectricity and ferroelasticity, coexist simultaneously in hole doped GdCl$_2$ monolayer (a ferromagnetic semiconductor). This is achieved by substituting 1/3rd of the Gd$^{2+}$ ions with Eu$^{2+}$ in the hexagonal structure of GdCl$_2$ monolayer. The resulting metallic state undergoes a bond-centered charge ordering driving a distortion in the hexagonal structure making it semiconducting again and {\it ferroelastic}. Further, the lattice distortion accompanied by a breaking of the lattice centrosymmetry renders a non-centrosymmetric charge distribution which makes the monolayer {\it ferroelectric}, at the same time. The two ferroic orders, ferroelectricity and ferroelasticity, present in Eu doped GdCl$_2$ monolayer are found to be strongly coupled making it a promising candidate for device applications. The doped monolayer remains a ferromagnetic semiconductor with large 4f magnetic moment just like the parent monolayer and possesses an even higher (out-of-plane) magnetic anisotropy energy (MAE) than its pristine counterpart as desired for two dimensional magnets to have high transition temperature.     
\end{abstract}

\maketitle

Multiferroic materials i.e. the materials possessing two or more ferroic orders simultaneously, present a potential avenue for multi-state non-volatile memory storage and various other multifunctional device applications via the cross coupling between different ferroic orders\cite{Eerenstein,Wang}. However, this coexistence of more than one ferroic orders is not very common, especially in two dimensional (2D) materials. It is due to the Mermin-Wagner theorem \cite{Mermin} which doesn't allow the spontaneous magnetic ordering in 2D materials and also, large depolarization field in ultra-thin films which destroys the vertical polarization\cite{Junquera,Fong}. Also, the mechanisms of ferroelectricity (FE) and ferromagnetism (FM) are mutually exclusive, thus making the coexistence of FE and FM and hence the realization of magnetoelectric multiferroics rare\cite{Spaldin}. Despite these difficulties, quite a few two-dimensional multiferroic materials showcasing two ferroic orders simultaneously have been predicted, some of which are also experimentally realized recently\cite{Tang,An}. However, only a very small number of 2D materials possessing all the three orders (FE, FM and ferroelastic (FA)) i.e \textit{triferroics} have been reported so far\cite{Yang,Shen, L. Yang}. These materials are very promising for multifunctional device purposes and hold possibilities for novel physical phenomena. Also, in order for these materials to be directly integrated into the conventional semiconductor circuits, it is preferable for them to be semiconducting in nature.

Recently, a series of 2D FM semiconducting monolayers based on rare-earth halides (GdX$_2$ (X=Cl, Br, I)) have been predicted\cite{B. Wang,Liu}. These monolayers have large magnetization: 8$\mu_B$ per formula unit (f.u.) owing to strong intra-ion Gd$_{4f}$-Gd$_{5d}$ orbital interaction and large magnetic anisotropy energies due to large spin-orbit coupling. Their Curie temperatures are predicted to be quite high (more than 220K) due to strong superexchange interaction between Gd-5d orbitals via halogen p-orbitals\cite{B. Wang, Liu}. The band gap values of monolayer GdCl$_2$, GdBr$_2$, and GdI$_2$ calculated within PBE+U method are found to be 0.89, 0.82 and 0.60 eV respectively\cite{Liu}. Recently, another rare-earth based monolayer GdI$_3$ with honeycomb lattice is identified as an antiferromagnet from first principles calculations which when doped with electrons via intercalation of metal atoms (Li and Mg) displays Fermi surface nesting\cite{You}. This makes the monolayer undergo Peierls transition which, assisted by the electron-phonon coupling, generates ferroelasticity in the monolayer. Simulataneouly, the ground state magnetic configuration of the monolayer changes from AFM to FM making it a FM-FA multiferroic.

In order to explore the possibilities for multiferroicity caused by similar phenomenon, we choose GdCl$_2$ monolayer, with the largest band gap among the GdX$_2$ (X=Cl, Br, I) monolayers, for hole doping. Although, we don't observe a similar Peierls dimerization as seen in GdI$_3$\cite{You}, we observe a bond-centered charge ordering within the unit cell which generates a spontaneous strain in the lattice. Owing to the three-fold rotational symmetry of the parent lattice, the charge ordering can take place along three different orientations which makes the strain switchable rendering the monolayer ferroelastic.  Also, the deformation of lattice leads to a breaking of centrosymmetry resulting in a finite spontaneous polarization. We further show that the ground state magnetic configuration of the monolayer remains ferromagnetic. Thus, the hole-doped GdCl$_2$ monolayer is found to be ferroelectric, ferromagnetic and ferroelastic simultaneously making it a single phase \textit{triferroic}. Also, the ferroelectricity and ferroelasticity of the monolayer are found to be strongly coupled.   

The first-principle calculations are performed using density functional theory (DFT) as implemented in the Vienna \textit{ab-initio} Simulation Package(VASP)\cite{Kresse}. In order to deal with the core electrons, we have used the projector augmented wave (PAW) method whereas the kinetic energy cut-off for the plane waves is set at 500 eV. For the exchange correlation functional, we have used the Perdew-Burke-Ernzerhof (PBE) parametrization\cite{Perdew} of the generalized-gradient approximation. The standard pseudopotentials were used for all the atoms with the elctronic configurations \(5s^25p^64f^75d^16s^2\) for Gd, \(5s^25p^64f^75d^06s^2\) for Eu and \(3s^23p^5\) for Cl atoms. The Hubbard correlation is taken into account using the rotationally invariant GGA (PBE) + U method \cite{Liechtenstein} with U$_{eff}$ = U - J = 4.0 eV (here U is Coulomb correlation and J is Hund's exchange) for both Gd and Eu 4f-orbitals, the value having been adopted from previous work\cite{Liu}. Treatment of f-electrons using PBE+U is possible as f-subshell is half filled. The structures are relaxed until the Hellmann-Feynman forces are reduced to 0.01 eV/$\AA$ and the energy convergence criteria is set at $10^{-6}$ eV. A $\Gamma$ centered $11\times11\times1$ Monkhorst-Pack grid was chosen for electronic structural optimization. There was a vacuum space of 15 $\AA$ between two layers to avoid interlayer interactions. The phonon dispersion calculations were performed using the Density Functional Perturbation Theory (DFPT) as implemented in VASP and the PHONOPY package with a $3\times3\times1$ supercell\cite{Gonze,Togo}. To check thermal stability, \textit{ab-initio} molecular dynamics (AIMD) calculations were performed where we used NVT ensemble at 300K using Nos{\'e} Hoover thermostat. The simulation was performed for 8.1 ps with a time step of 3 fs\cite{Martyna}. 
\begin{figure}[!t]
\centering
\includegraphics[width=9cm]{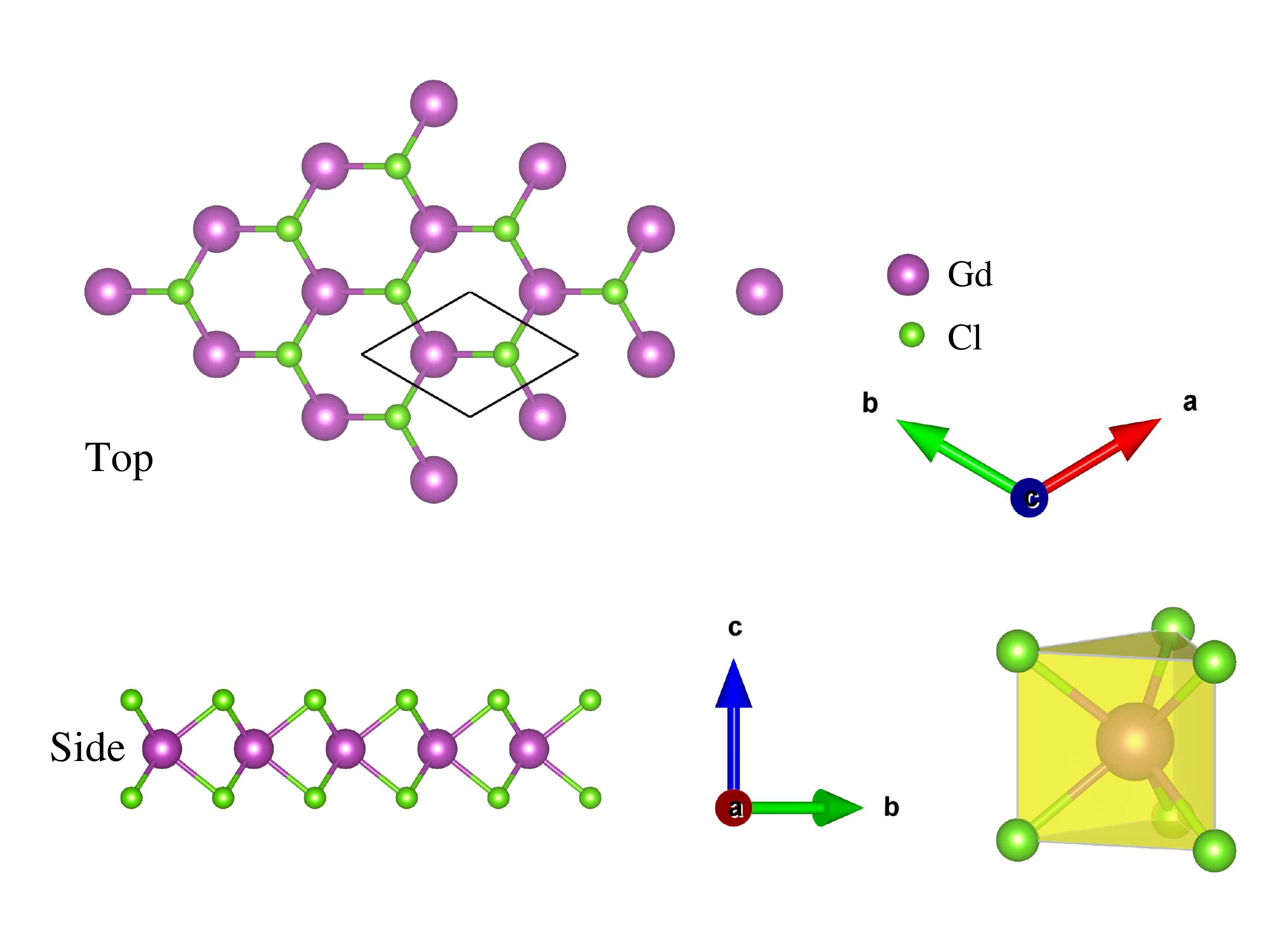}
\caption{Top (upper panel) and side (lower panel) views of the crystal structure of GdCl$_2$ monolayer where solid black rhombus in the top view represents the primitive cell. GdCl$_6$ trigonal prismatic cage is shown in yellow on the right side of the lower pannel.}
\end{figure}

GdCl$_2$ monolayer has hexagonal crystal structure with space group $p\bar{6}m2$. Each monolayer consists of three atomic layers : one Gd atom layer sandwiched by two Cl atom layers as shown in Fig.1. Each Gd ion is coordinated by 6 Cl atoms forming a trigonal prismatic cage (see Fig.1). The calculated optimized lattice constants of GdCl$_2$ monolayer are $a=b=3.78437 \AA $ which are similar to the previously reported values\cite{Liu}. Also, it is found to be a FM semiconductor with a magnetic moment of 8$\mu_B$ per Gd atom and a band gap of 0.88 eV, in close agreement with the reported values\cite{Liu}.

\begin{figure}[!b]
\centering
 \includegraphics[width=9.0cm]{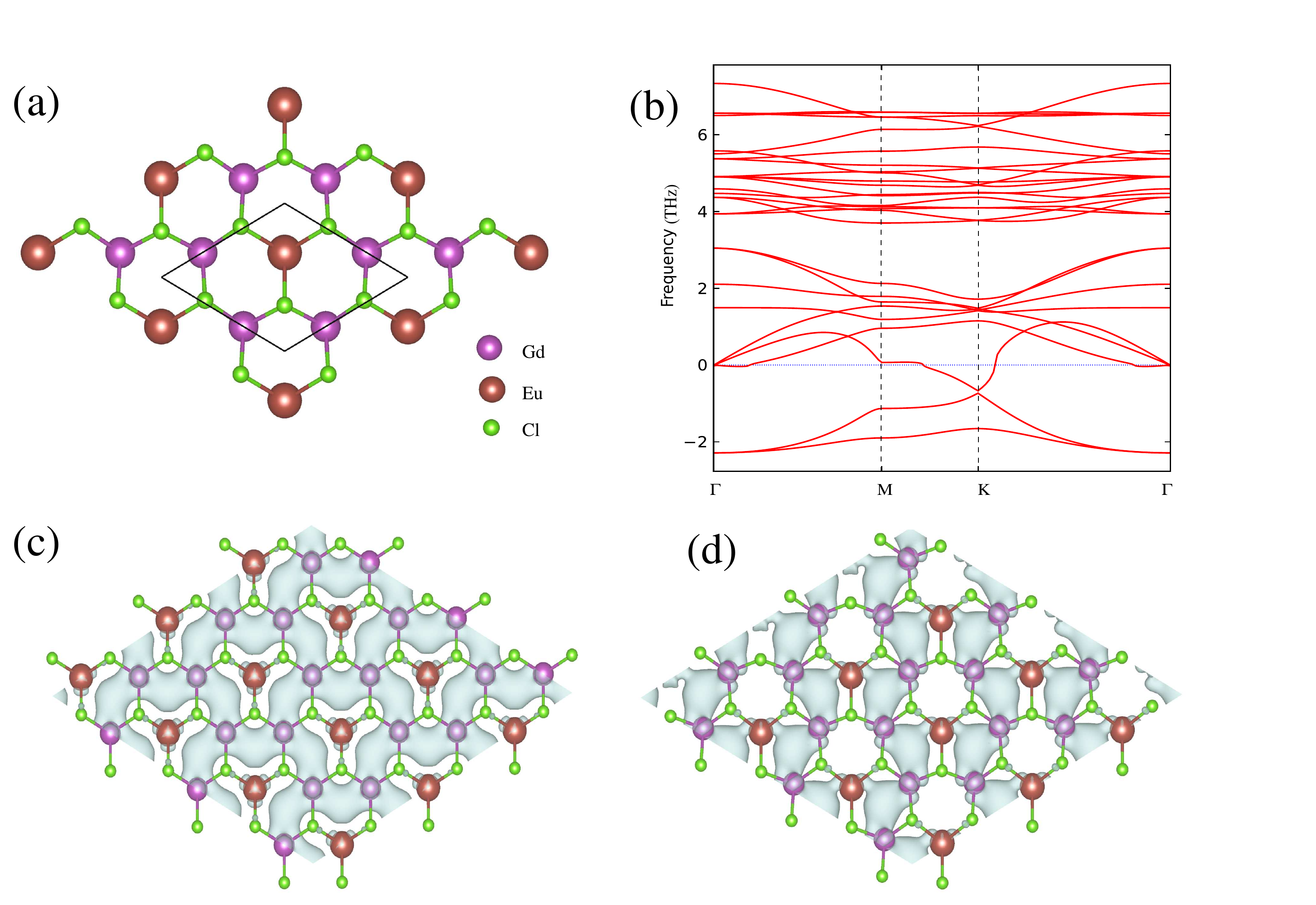}
 \caption{(a)Top view of high-symmetry hexagonal 1/3 Eu-doped GdCl$_2$ monolayer. (b) Phonon dispersion for the hexagonal symmetric structure of doped monolayer showing soft phonon modes. (c) Centrosymmetric and (d) non-centrosymmetric charge density distributions before and after relaxation of the doped strcture respectively.}
\end{figure}

\begin{figure*}[!ht]
\centering
 \includegraphics[width=16.0cm]{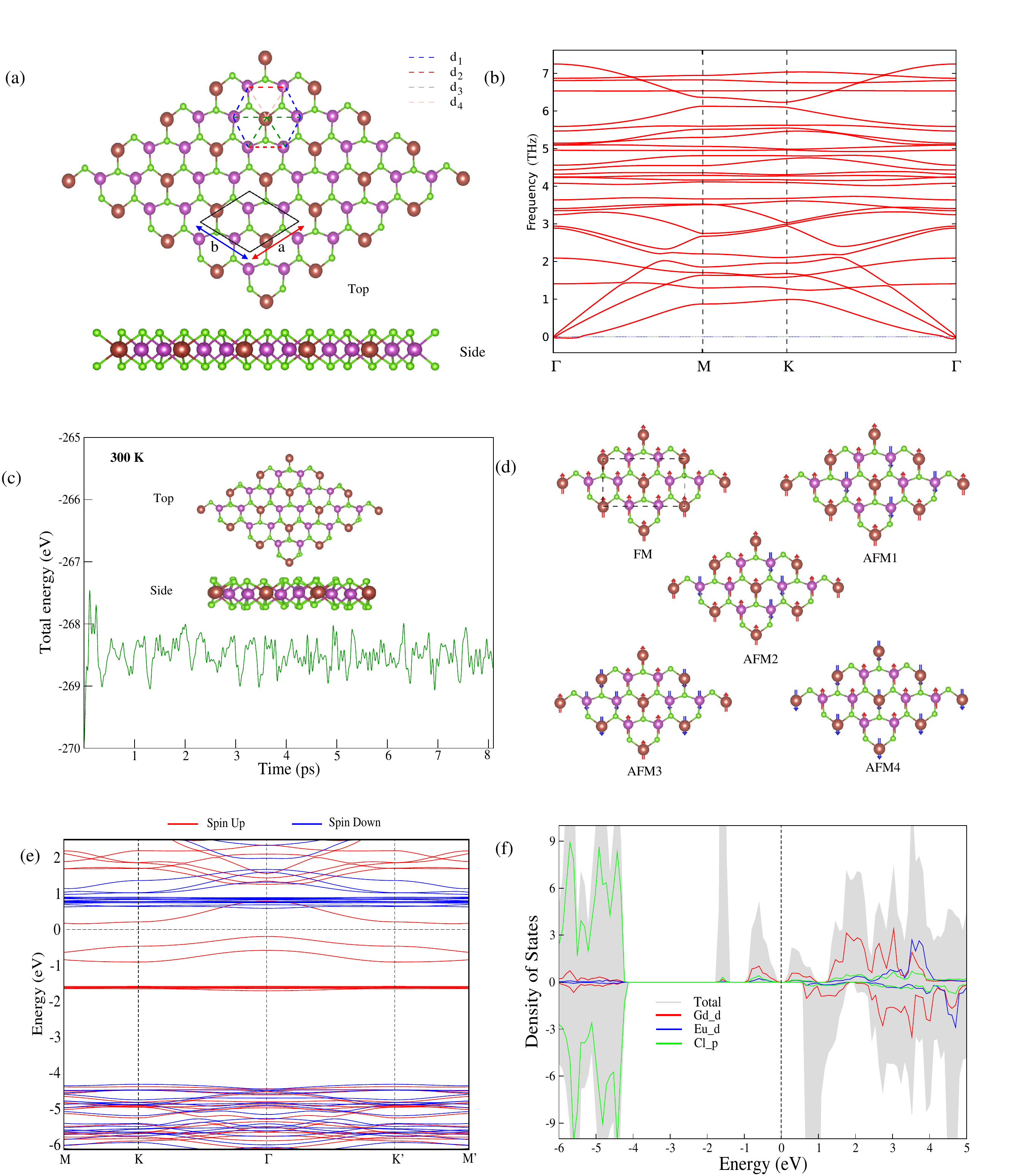}
 \caption{(a) Top and side views of the Eu-doped GdCl$_2$ monolayer after relaxation. Red (Blue) lines represent long (short) NN Gd-Gd bonds and brown (green) lines represent long (short) NN Gd-Eu bonds.(b) Phonon dispersion spectrum for the relaxed structure (c) The fluctuation of total energy and snapshots (top and side view) of monolayer structure after 8.1 ps from AIMD simulation. (d) Various possible magnetic configurations for doped monolayer. Red and Blue arrows represent spin up and spin-down respectively. Black rectangle represents the unit cell used for magnetic energy calculations. (e) Spin resolved band structure and (f) Total and orbital-projected partial density of states (PDOS) of the deformed doped monolayer.}
\end{figure*}
To explore the possibility of making the monolayer ferroelctric, we doped $1/3$rd of all Gd sites by Eu in GdCl$_2$ monolayer i.e we substituted Gd atoms present at the center of each Gd hexatomic ring with Eu atoms (Fig. 2(a)). The formation energy for this doping is found to be -6.71 eV per formula unit as calculated using the formula $$ E_f = E(Gd_2EuCl_6) - E(Gd_3Cl_6) + \mu_{Gd} - \mu_{Eu} $$ where E(Gd$_2$EuCl$_6$) and E(Gd$_3$Cl$_6$) represent the energy of Eu-doped and pristine $ \sqrt{3}\times\sqrt{3}\times1$ supercell of GdCl$_2$ monolayer respectively whereas $\mu_{Gd}$ and $\mu_{Eu}$ represent the chemical potentials of Gd and Eu. We take energy per atom of the bulk Eu (orthorhombic) as $\mu_{Eu}$ and the difference in energy of GdCl$_2$ ML unit cell and the Cl$_2$ molecule as $\mu_{Gd}$. The negative value of the formation energy confirms the feasibility of the doping process. The oxidation state of Eu atoms doped in GdCl$_2$ ML is +2, same as that of Gd and hence, their electronic configuration is \(4f^7d^0\). The Eu d-orbitals are energetically close to the Gd d-orbitals causing a finite overlap between them. Thus, the given Eu doping amounts to a 1/3rd hole doping in the valence band of the pristine GdCl$_2$ monolayer making the doped monolayer metallic. However, the resulting metallic hexagonal structure of the doped monolayer is unstable, as indicated by the soft phonons present in its phonon spectrum (see Fig. 2(b)), towards a bond centered charge ordered state accompanied by the lattice distortions where the charge density localizes on four out of six Gd-Gd bonds on the hexagonal ring of Gd ions surrounding each Eu (Fig. 2(d)). For comparison we show in Fig. 2(c) the uniform charge density distribution on all the six Gd-Gd bonds in the unstable structure. The corresponding lattice distortions are obtained upon performing a full relaxation of the unit cell (removing all the symmetries present). The four Gd-Gd bonds with finite charge density residing on them are found to be shortened while the other two are elongated. Also, the Eu atoms initially placed at the center of the Gd hexatomic rings shift away from the center making the structure non-centrosymmetric. The final relaxed structure is thus triclinic with space group $p1$. The top and side views of the relaxed structure are shown in Fig. 3(a) with various Gd-Gd/Gd-Eu bond lengths (d1-d4) marked. The comparison between the nearest neighbour Gd-Gd and Gd-Eu bond lengths and the lattice parameters in the initial hexagonal and the relaxed deformed structure is given in Table I.  The phonon spectrum for the relaxed structure no longer shows the soft phonons indicating its dynamical stability (Fig. 3(b)). Further, to check the thermal stability, we performed the \textit{Ab-initio} Molecular Dynamics (AIMD) simulation at temperature 300K for 8.1 ps with a time step of 3 fs. There are only small fluctuations in energy and the structure remains almost intact after 8.1 ps (Fig. 3(c)). 

In order to understand the bond centered charge ordering and associated lattice distortions observed in our system, we looked at various possible scenarios. Though we observed Fermi surface nesting in the high symmetry metallic phase, we do not observe any unit cell doubling expected from Peierls dimerization as observed in GdI$_3$\cite{You}. So we believe the charge ordered state that we observe in our system is unlikely to be associated with Peierls-like transition. This charge ordered state could be similar to the ``Zener polaron order" state observed in doped manganites\cite{Aladine,Zhou} which involves both Zener double exchange and a polaronic-like distortion between the Gd and Eu sites with finite charge density between them. As Zener double exchange is known to drive a ferromagnetic order between localized spins and also as all the Gd and Eu atoms are shared by more than one such `Zener polaron bonds', the spins at all the atoms in the crystal are expected to align in a FM pattern. However, further calculations and experiments are needed to establish the mechanism. 

To probe the magnetic ground state of Eu-doped GdCl$_2$ ML, we studied four different AFM configurations alongwith the FM one (Fig. 3(d)). The preferred magnetic ordering is found to be FM with AFM1, AFM2, AFM3 and AFM4 magnetic configurations found to be 575.0 meV, 208.9 meV, 563.5 meV and 378.9 meV per supercell (shown in Fig. 3(d)) higher in energy respectively. The value of magnetization per formula unit (Gd$_2$EuCl$_6$) is found to be 21.76$\mu_B$ which is in close agreement with the formal magnetic moment of 8$\mu_B$ on Gd with electronic configuration \(4f^7d^1\) and 7$\mu_B$ on Eu with electronic configuration \(4f^7d^0\), in valence state +2. In Fig. 3 (e) and (f) we present the band structure and density of states respectively for FM state. As evident from these figures the low symmetry doped monolayer is a semiconductor with an indirect band gap of about 0.40 eV. The bands above and below the Fermi level (FL) mainly comprise of spin up Gd d-orbitals with only a small contribution from spin up Eu d-orbitals. The Cl 2p-orbitals mainly lie far below the FL with a little overlap of Gd and Eu d-orbitals. Also, the localized 4f orbitals lie far below the FL. Thus, Eu-doped GdCl$_2$ ML is a half semiconductor with only spin up states closest to the FL. In order to determine the preferred direction for spin-alignment, we performed spin-orbit coupling calculation which is quite important in our case as we are dealing with heavy rare-earth ions. The preferred spin orientation is found to be out of plane of the monolayer. The magnetic anisotropy energy (MAE) defined as the energy difference between the in-plane and out of plane spin orientations i.e. $\Delta E=E_c-E_{ab}$ is found to be about 790$ \mu $eV/f.u. which is higher than that of pristine GdCl$_2$ monolayer\cite{Liu} and hence can play a big role in stabilizing the magnetic ordering in the 2D limit. The band structure doesn't change much after including SOC and the band gap also remains almost unchanged. 

In order to get an estimate of the value of \(T_c\) of the doped monolayer, we calculate the nearest neighbour magnetic exchange interaction parameters. In the deformed structure of the monolayer, we have four different kinds of nearest neighbour pairs: short and long distance Gd-Gd pairs, distances denoted by \(d_1\) and \(d_2\) respectively, and short and long distance Gd-Eu pairs, distances denoted by \(d_3\) and \(d_4\) respectively (see Fig. 3(a) and Table I). Correspondingly, there are four kinds of NN exchange parameters denoted by \(J_1\), \(J_2\), \(J_3\) and  \(J_4\). In order to calculate these parameters, we use the energy mapping scheme\cite{Liu} using the spin Hamiltonian: \[H=-J\sum_{<i,j>}S_i.S_j\] where J is the nearest neighbour exchange parameter, S$_i$ and S$_j$ are spin operators. We use five different magnetic configurations as illustrated in Fig. 3(d). In accordance with the relative spin alignment of the nearest neighbouring atoms, the magnetic energy expressions for these configurations per magnetic unit cell (supercell) shown in Fig. 3(d) are given as:  
\begin{align}
E_1&=E_0-8J_1|S_a|^2-4J_2|S_a|^2-16J_3S_a.S_b-8J_4S_a.S_b\\
E_2&=E_0+8J_1|S_a|^2+4J_2|S_a|^2\\
E_3&=E_0-8J_1|S_a|^2+4J_2|S_a|^2\\
E_4&=E_0+8J_1|S_a|^2-4J_2|S_a|^2-8J_4S_a.S_b\\
E_5&=E_0-8J_1|S_a|^2-4J_2|S_a|^2+16J_3S_a.S_b+8J_4S_a.S_b
\end{align}

where E$_0$ represents the energy without magnetic coupling whereas S$_a$ and S$_b$ represent the spins of Gd and Eu atoms respectively determined by the number of unpaired electrons in valence state +2. The values of exchange parameters calculated using these expressions are found to be J$_1$ =1.43 meV, J$_2$=0.173 meV, J$_3$=-0.083 meV, J$_4$=0.887 meV. Thus, the exchange parameters corresponding to shorter Gd-Gd and Gd-Eu bonds are significantly larger than the ones corresponding to longer bonds. The larger values of the exchange parameters are comparable to those of the pristine GdCl$_2$ monolayer where Monte-Carlo simulations show a T$_c$ of about 224K\cite{Liu}. We therefore expect our system also to support a comparable T$_c$ .

\begin{table}[!h]
\centering
\renewcommand{\arraystretch}{1.5}
\label{tab1}
\caption{Nearest neighbouring Gd-Gd and Gd-Eu pair distances and the lattice parameter comparison between the high-symmetry ($p\bar{6}m2$) and the relaxed deformed ($p1$) structure of Gd$_2$EuCl$_6$ monolayer.}

\scalebox{1.1}{
\begin{tabular}{|c|c|c|c|c|c|c|}
\hline
\textbf{Structure} & \textbf{$d_1(\AA)$} & \textbf{$d_2(\AA)$} & \textbf{$d_3(\AA)$} & \textbf{$d_4(\AA)$} & \textbf{$a(\AA)$} & \textbf{$b(\AA)$} \\
\hline
Hexagonal & 3.93 & 3.93 & 3.93 & 3.93 & 6.80 & 6.80 \\
\hline
Deformed & 4.30 & 3.78 & 4.23 & 3.82 & 6.85 & 6.79 \\
\hline
\end{tabular}}
\end{table}
\begin{figure}[!b]
\centering
 \includegraphics[width=9.0cm]{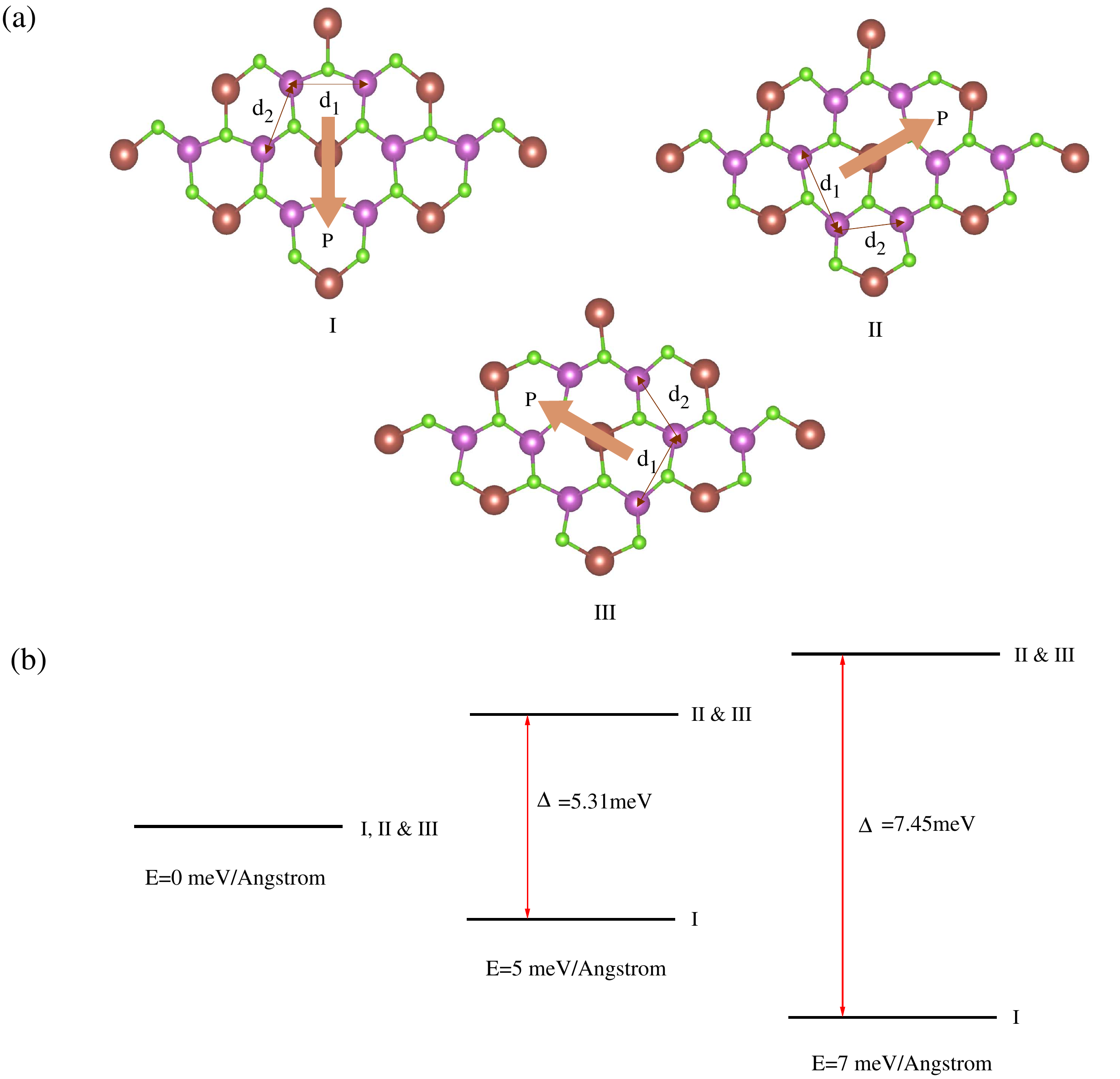}
 \caption{(a)Different possible lattice configurations of the doped monolayer and corresponding spontaneous polarization directions. (b)Energies of different lattice configurations under external electric field applied along polarization direction of lattice configuration I.}
\end{figure}

\begin{figure}[!t]
\centering
\includegraphics[width=8.5cm]{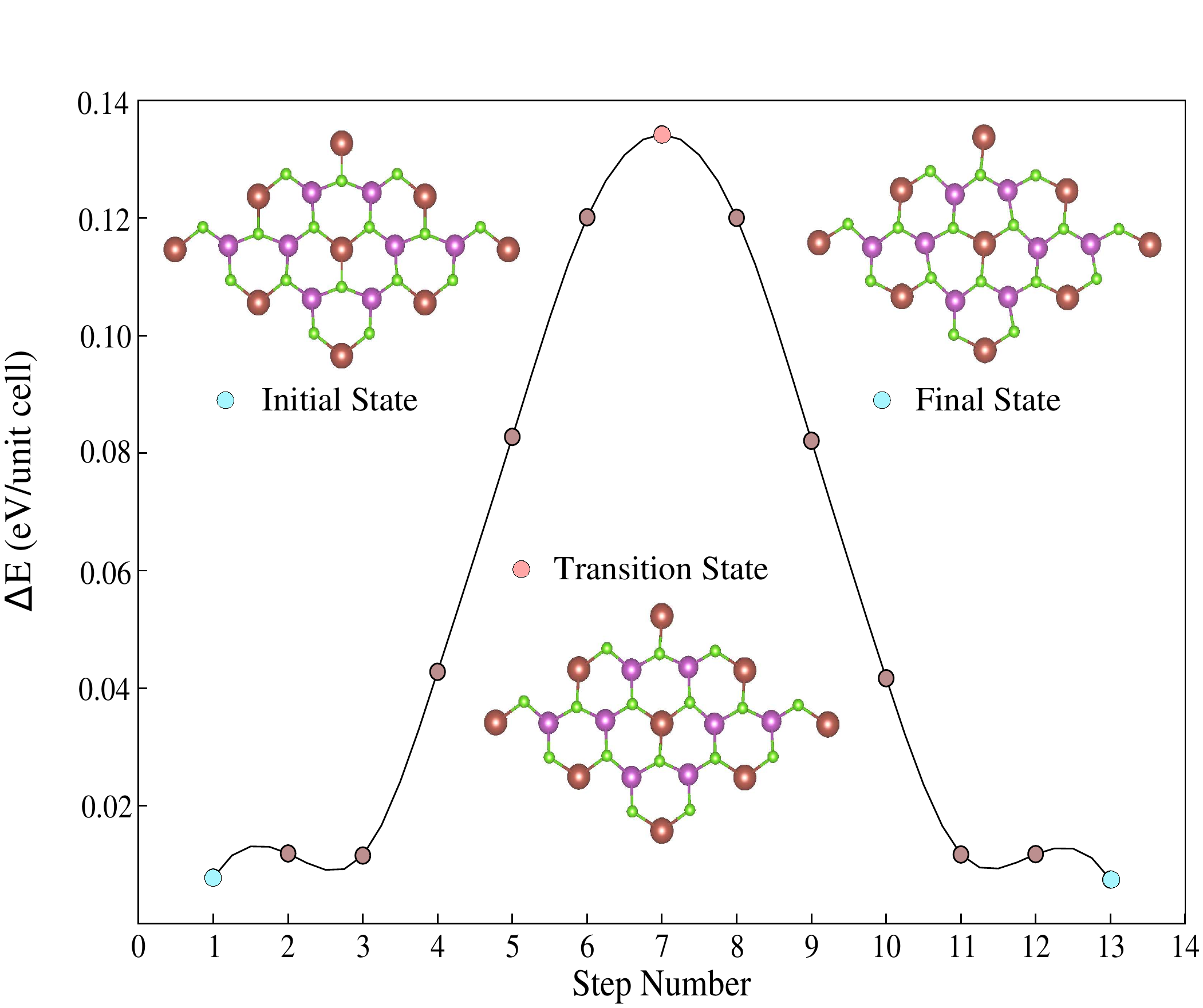}
\caption{Minimum energy pathway for transition from one ferroelastic(ferroelectric) phase to another as obtained using Solid State Nudged Elastic Band (SS-NEB) method. In inset, are the initial, final and the highest energy state acquired by the monolayer during the structural transition.}
\end{figure}

As a result of the bond centered charge ordering associated lattice distortion, the three-fold rotational symmetry of the initial hexagonal structure is broken. The deformed structure has an increased lattice constant along $a$-direction (6.85 $\AA$) and a shrunk lattice constant along $b$-direction (6.79$\AA$) (Fig. 3(a)) i.e. $a$ is around 1\% larger than $b$. Now, as the parent high-symmetry structure is hexagonal and has three-fold rotational symmetry, this deformation of the lattice can occur along three different orientations (Fig. 4(a)) all of which are equally probable. Thus, the Eu-doped GdCl$_2$ monolayer can relax into three equivalent lattice configurations oriented at 120$^\circ$ with respect to each other. As a result, Eu-doped GdCl$_2$ ML manifests ferroelasticity where one can switch between three equivalent lattice configurations by changing the strain direction via external mechanical stress. This switching between different configurations in the presence of external forces also makes the material stable against mechanical damage.  

Interestingly, since the deformed structure also involves an off-center shifting of Eu atoms leading to a non-centrosymmetric charge density distribution, we observe a net spontaneous polarization, estimated from our calculations to be 170 pC/m, along the direction of Eu shifting (Fig. 4(a)). The magnitude of polarization is more than that of ML Bi$_2$O$_2$S (12 pC/m) \cite{Wu} and  (CrBr$_3$)$_2$Li (92 pC/m) \cite{Huang} but less than that of ML GaTeCl (578 pC/m) \cite{S.-H. Zhang}. Since there are three possible directions for Eu off-center shifting due to the three-fold rotational symmetry of the initial hexagonal lattice, one can have, correspondingly, three different polarization directions which point at an angle of 120$^\circ$ with respect to each other as shown in Fig. 4(a). With the help of an external field, it should be possible to switch between different polarization directions making the Eu-doped GdCl$_2$ monolayer ferroelectric along with being ferroelastic and ferromagnetic. Thus, the Eu-doped GdCl$_2$ ML can be termed as a single-phase triferroic. In order to calculate the value of spontaneous polarization, we used modern theory of polarization as implemented in VASP\cite{Resta,Smith}. The high symmetry non-polar lattice configurations are, however, found to be metallic and hence, their polarization cannot be calculated using the Berry phase approach. Therefore, in order to remove the possibility of any polarization quanta, we made sure that the polarization orients in the same direction as that of Eu shifting. Also, we calculated the polarization values for two different lattice configurations at 120$^\circ$ relative orientation and ensured that they had the same magnitude but were directed at 120$^\circ$ w.r.t. one another.

As the direction of spontaneous polarization is decided by the direction of lattice deformation and associated charge ordering, the ferroelectricity and ferroelasticity are strongly coupled. Thus, the ferroelectric switching can be carried out by switching between lattice configurations with different charge ordering directions (different strain directions) using an external mechanical stress. Further, it can be shown that on applying an external electric field, the lattice configuration with polarization direction along the direction of electric field becomes energetically favourable. For this purpose, we applied an in-plane external homogeneous electric field along the polarization directions in the lattice configurations I, II and III using the Perturbation Expression After Discretization (PEAD)\cite{Ding,Nunes} approach as implemented in VASP. We observe that when the electric field is applied along the polarization direction of lattice configuration I, the energy of lattice configuration I decreases while that of the configurations II and III increase (Fig. 4(b)). Same can be shown for the external electric field along the polarization directions of other configurations too. Thus, depending on the direction of an applied external electric field, the monolayer may undergo a structural transition involving a change in the direction of polarization and also of the lattice strain. Therefore, an external electric field can be used as a handle to switch ferroelasticity(the strain direction) in the system along with ferroelectricity(polarization). 

In order to investigate the energetics of the structural transition from one ferroelastic(ferroelectric) state to another, we performed a solid-state nudged elastic band (SS-NEB) calculation. The minimum energy pathway (MEP) for transition from one lattice configuration to another rotated at 120$^\circ$ is shown in Fig. 5. The transition barrier height is 120 meV/f.u. which is more than that of 2D FE-FA multiferroic MX monolayers (M=Sn, Ge; X=S, Se)\cite{H. Wang} but less than that of GaTeCl monolayer\cite{S.-H. Zhang}. Such an energy barrier indicates that the FE-FA phases of Eu-doped monolayer are robust at room temperature but still can be made to transit from one to the other by application of a moderate amount of external mechanical stress or electric field.

In summary, we predict based on a detailed theoretical investigation that Eu doped GdCl$_2$ monolayer would be a promising candidate material for the realization of two-dimensional single phase triferroic where ferromagnetic, ferroelectric and ferroelastic orders coexist. The doped monolayer is found to be thermodynamically stable from our phonon and AIMD calculations. The estimation of magneto-anisotropy energy and exchange parameters indicate that the ferromagnetic state can have a comparable transition temperature to that predicted for the parent compund GdCl$_2$ ($\sim$220K). The hole doped GdCl$_2$ monolayer undergoes a bond centered charge ordering accompanied by lattice distortions which opens a gap at the Fermi surface making the system semiconducting. In addition we observe shifting of Eu ions from centrosymmetric positions giving rise to ferroelectricity in the system. We have also shown through our calculations that there exists a strong coupling between ferroelectricity and ferroelasticity which can be switched among three possible orientations via an external electric field or mechanical stress. Thus, Eu doped GdCl$_2$ monolayer looks very versatile material for device applications.           

TM acknowledeges discussions with P. K. Choubey. SB acknowledges Council of Scientific and Industrial Research (CSIR), India for research fellowship. TM acknowledges Science and Engineering Research Board (SERB), India for funding support through MATRICS research grant (MTR/2020/000419). TM and SB acknowledge PARAM GANGA computational facility at IIT Roorkee.

\end{document}